\newif\ifmnras
\mnrastrue
\ifmnras
    \documentclass[apj,numberedappendix]{emulateapj}
\else
	\documentclass[apj,numberedappendix]{emulateapj}
\fi

\usepackage{graphics,epsf}
\usepackage{amsmath}                
\usepackage{amsfonts}               
\usepackage{amssymb}                
\usepackage{epsfig}                 
\usepackage{graphicx}
\usepackage{float}
\usepackage{color}

\def \s{~\rm{s}}
\def \km{~\rm{km}}

\def \AU{~\rm{AU}}

\def \yr{~\rm{yr}}

\ifmnras
	\def \aap{A\&A}
	
	\def \apj{ApJ}
	\def \apjl{ApJ}

	\def \mnras{MNRAS}

\fi

\usepackage{xcolor}
\definecolor{redak}{rgb}{0.9,0.15,0.05}

\begin{document}

\ifmnras

\title{Oxygen-neon rich merger during common envelope evolution}
\author{Pere Canals \altaffilmark{1}, Santiago Torres\altaffilmark{1,2} \& Noam Soker\altaffilmark{3,4}}

\altaffiltext{1}{Departament de F\'\i sica, Universitat Polit\`ecnica de Catalunya, c/Esteve Terrades 5, 08860 Castelldefels, Spain; perecanals.7@gmail.com}
\altaffiltext{2}{Institute for Space Studies of Catalonia, c/Gran Capit\`a 2--4, Edif. Nexus 104, 08034 Barcelona, Spain; Santiago.Torres@upc.edu}
\altaffiltext{3}{Department of Physics, Technion -- Israel Institute of Technology, Haifa 32000, Israel; soker@physics.technion.ac.il}
\altaffiltext{4}{Guangdong Technion Israel Institute of Technology, Shantou 515069, Guangdong Province, China}

\begin{abstract}
We conduct a population synthesis study of the common envelope evolution (CEE) of a white dwarf (WD) and an asymptotic giant branch (AGB) star and find that the potential number of type Ia supernovae (SNe Ia) from the core degenerate (CD) and from the double degenerate (DD) scenarios are of the same order of magnitude.
For the CD scenario we consider cases where a carbon oxygen rich (CO) AGB core and a CO WD merge during the CEE and leave a WD remnant with a mass of $M_{\rm WD} \ge 1.35 M_\odot$, and for the DD scenario we count surviving CO WD binary systems that merge within a time of $10^{10} \yr$.
When either the AGB core or the WD are oxygen neon rich (ONe) we assume that the outcome might be a peculiar SN Ia. We find that the number of potential peculiar SNe Ia in the channels we study, that do not include peculiar SNe Ia that involve helium WDs or helium donors, is non-negligible, but less than the number of normal SNe Ia.
If a SN Ia or a peculiar SN Ia explosion takes place within about million year after CEE, whether in the CD scenario or in the DD scenario, a massive circumstellar matter is present at explosion time.
Our results are compatible with the suggestion that Chandrasekhar-mass SNe Ia mostly come from the CD scenario, and sub-Chandrasekhar SNe Ia mostly come from the DD scenario.
\end{abstract}



\section{INTRODUCTION}
\label{sec:intro}

Observations and theoretical studies in recent years strengthen the notion that in some cases a white dwarf (WD) that spirals-in inside the envelope of a giant star, e.g., a red giant branch (RGB) or an asymptotic giant branch (AGB) star, merges with the giant's core.
One type of observations that suggest the merger process of a WD with the core of a giant is the bias of WDs with strong magnetic fields to be more massive than the general WD population; some of these highly magnetic WDs can result from a core-WD merger  (e.g.,  \citealt{Toutetal2008, Briggsetal2018}).
A hint for possible merger comes also from the new finding from {\it Gaia} observations. The Hertzsprung-Russell diagram obtained for the {\it Gaia}-DR2 WD population shows a clear bifurcation which implies a bimodal-like mass distribution with an excess of WDs with masses of about $0.8 M_\odot$ \citep{Jimenezetal2018, Kilicetal2018}. This bifurcation can not be explained by recent models of single WD evolution, neither invoking differences in WD atmospheric composition. Although the actual reasons of the bifurcation remain unclear, one plausible explanation is the contribution of WD mergers \citep{Jimenezetal2018, Kilicetal2018}.

Another motivation to consider the merger of WDs with CO cores of AGB stars is the quest for the progenitors of type Ia supernovae (SNe Ia). There are five different binary scenarios to bring a WD or two WDs to explode as a SN Ia: the core degenerate (CD) scenario, the double degenerate (DD) scenario, the double-detonation scenario, the single degenerate scenario, and the WD-WD collision scenario (for recent reviews that list these 5 scenarios, see, e.g., \citealt{LivioMazzali2018, Soker2018Rev, Wang2018}; and references to many earlier papers and reviews therein). In the present paper we will refer only to the CD and DD scenarios.

In the CD scenario a CO WD merges with the CO core of an AGB star at the final stages of the common envelope evolution (CEE), and forms a CO WD with a mass of about the Chandrasekhar mass $M_{\rm WD} \simeq M_{\rm Ch} \simeq 1.4 M_\odot$.
Studies during the last decade have developed the CD scenario as a distinguished SN Ia scenario (e.g., \citealt{KashiSoker2011, IlkovSoker2013, AznarSiguanetal2015}), and argue that it might account for most SNe Ia.

In the DD scenario the WD and the core survive the CEE, and they merge a long time after the CEE has ended (e.g., \citealt{Webbink1984, IbenTutukov1984}). The time from the merger itself until explosion takes place (the merger explosion delay, or MED) that is required and allowed by the DD scenario is still an open question (e.g., \citealt{LorenAguilar2009, vanKerkwijk2010, Pakmor2013, Levanonetal2015}; see review by \citealt{Soker2018Rev}).

Another useful classification is to SN Ia scenarios where the WDs explode with masses near the Chandrasekhar mass limit, `$M_{\rm Ch}$ explosions', including the CD and the single-degenerate scenarios, and scenarios where WDs explode with lower masses, `sub-$M_{\rm Ch}$ explosions' (e.g., \citealt{Maguireetal2018}), including the DD, the double-detonation, and the WD-WD collision scenarios.
As at least some SNe Ia are $M_{\rm Ch}$ explosions (e.g., \citealt{Ashalletal2018} for a recent study) the DD scenario is unlikely to account for all SNe Ia.
The studies by \cite{Dhawanetal2018} and \cite{Diamondetal2018} suggest that SN~2014J was an $M_{\rm Ch}$ explosion, probably the CD scenario \citep{Soker2015}.
In another recent study \cite{BearSoker2018} find that there are sufficient number of cataloged massive WDs, $M_{\rm WD} > 1.35M_\odot$, that might potentially explode as SNe Ia in the frame of the CD scenario. Further support to the CD scenario comes from the finding of \cite{Cikotaetal2017} that some proto-planetary nebulae have polarization curves similar to those observed in some SNe Ia.

In light of the potential of the CD scenario to account for many, or even most, SNe Ia, in the present study we consider variants of that scenario. Instead of the merger of a CO WD with a CO core, we consider a merger of a CO WD with an ONe core of an AGB star, or the merger of an ONe WD with a CO AGB core. Such rare mergers might lead to a peculiar SN Ia \citep{Kashyapetal2018}, or remnants that eventually will experience peculiar SN Ia explosions.

We study the formation of such rare mergers with a population synthesis code.
We used this code in an earlier paper where we study the CD scenario for the SN Ia PTF~11kx \citep{Soker2013}, where there is a massive circumstellar medium \citep{Dilday2012}. We describe this code in section \ref{sec:code}. In section \ref{sec:numbers} we present the relative numbers of systems formed in seven evolutionary channels, and in section \ref{sec:properties} we describe some properties of the binary systems that merge during the CEE.  We summarize in section \ref{sec:summary}.

\section{Population synthesis code}
\label{sec:code}
\subsection{Initial parameters}
\label{subsec:popsyn}

Following the work of \cite{Soker2013}, we use an updated version of the population synthesis code described in \cite{GBerro2012}, first presented in \cite{Camacho2014} and further developed by \cite{Cojocaru2017}. For the sake  of conciseness we present here the most important inputs. We generate  an ensemble of stellar systems with a constant star formation rate  during 10 Gyr, taken to be the approximate age of the disk of our Galaxy. Star masses are generated according to a Salpeter's Initial Mass Function (IMF)   with a canonical slope parameter $\alpha=-2.35$ \citep{Salpeter1955} and a metallicity value of Z=0.014. The mass range used,  0.4 M$_{\odot}\le\, M\,\le$ 50 M$_{\odot}$, is the most suitable for the gross of stars that are of interest to our work. As we deal only with binary stars that do not end up as helium WDs, the lower limit has no influence on our results as long as it is below about $0.8 M_\odot$.

As we only examine the relative fraction of different binary routes, the fraction of binary systems among all stars has no influence on the present study. The initial mass ratio distribution (IMRD) $n(q)$, i.e., the fraction of binaries as function of the mass ratio $q \equiv M_2/ M_1$, follows a flat distribution as assumed by \cite{Cojocaru2017}. The initial semi-major axis $a$ between the two stars forming the binary system is obtained through a logarithmically flat distribution in the range  $3 \le a/R_{\odot} \le 10^{6}$, usually applied for this kind of simulations (e.g., \citealt{IbenTutukov1984, Nelemans2001, Hurley2002a, Davis2010, Cojocaru2017}). The eccentricity $e$ of the binary is selected randomly for each system according to a thermal distribution, $f(e)=2e$ with $0\le e<1$ \citep{Heggie1975}.

\subsection{Binary systems treatment}
\label{subsec:binevo}

We use the binary stellar evolution code BSE \citep{Hurley2000} to evolve the binary systems we generate in our population synthesis model.
To describe the CEE we adopt the alpha-formalism (e.g., \citealt{Tout1997,DeMarco2011}). Although the CEE is not well understood and several modelings exist, we consider the alpha-formalism as the most standard approximation. The alpha-formalism basically assumes energy conservation of orbital and envelope-binding energy, and it parametrizes the energy transferred from the binary system to the envelope by a set of parameters. The most relevant parameter in the formalism is $\alpha_{\rm CE}$, which accounts for the efficiency of the system to convert orbital energy into energy to expel the envelope. There is no general agreement about the best choice for $\alpha_{\rm CE}$,  but a value in the interval 0.1-0.5 is generally adopted (e.g.,  \citealt{Zorotovic2014, Camacho2014}). Additionally, we also take into account another parameter $\alpha_{\rm int}$ (sometimes found as $\alpha_{\rm th}$) which accounts for the efficiency by which the internal energy of the system, including basically thermal energy, recombination and radiation, can be converted into kinetic energy to expel the envelope \citep{Han1995}. For our simulations we choose a value of $\alpha_{\rm int}$ between 0 and 0.1, following the criteria of \cite{Camacho2014}.

We focus on those systems that end up in a merger between a core of a giant star and a WD during the CEE. Additionally, we choose both the core and the WD to be more massive than 0.5 M$_{\odot}$, as we are  looking at CO/ONe rich degenerate remnants, and we consider the cores and WDs with a mass below this value to be mainly He-rich WDs. We assume that WDs more massive than 1.1 M$_{\odot}$ correspond to stars that achieve C-burning, leading to an enrichment of Ne and Mg \citep{GBerro1994,Althaus2007}. Specifically, WDs with masses above $\sim 1.05 M_{\odot}$ are supposed to be carbon-oxygen-neon (CONe) hybrid WDs, but since we are not going to consider these WDs in the simulations as a separate category, we simplify the criterion. In short, we  assume that any WD with a mass above 1.1 M$_{\odot}$ is already an ONe rich object, as usually adopted in these type of works (e.g., \citealt{Camacho2014, Meng2014}).

To determine whether a merger has taken place we follow the evolution of the stellar type of both stars. As we are looking for systems that merge with one of the two stars being a WD and the other being a giant star, we first identify those systems for which their final evolution is a single WD star with no companion. Afterwards, we derive the instant when the merger happens and we then study the progenitors of such events. For these systems, we derive the mass of the WD ($M_{\rm WD}$), the mass of the core of the giant star ($M_{\rm Core}$), the mass of the envelope of the giant star ($M_{\rm env}$), the initial semi-major axis of the binary system ($a$) and the stellar type of the post-merger remnant.

One of our main channels of interest consists in those systems in which the mass of the giant's core at the time of the merger is within the interval of $1.6 M_{\odot} \le M_{\rm Core}\le 2.25 M_{\odot}$. The reason behind this resides in that these stars could have massive enough cores to burn C \citep{Hurley2000} and thus lead to ONe-rich mergers during the CEE. This mass accounts for the CO/ONe core plus the He burning layer enclosing the inner denser core. In this range, the inner CO/ONe core would have a mass in the range $1.08 M_{\odot} \le M_{\rm InCore} \le M_{\rm Ch}$. These systems are particularly interesting because they lead to very massive WDs, even super-Chandrasekhar WDs, with ONe from the giant's core, that would have delayed explosions due to a very fast rotation after the merger. It is also worth noting here that these systems may either not explode, or explode as a peculiar SN Ia due to the large fraction of ONe in the core.

\subsection{Close double degenerate systems}
\label{subsec:DD}

Apart from these systems that end the CEE in merger, we also look at cases that end with a close binary system of two WDs, the WD and the former core of the AGB star. If the orbital separation is close enough they merge during a Hubble time, and if both are CO WDs, the merger might lead to a SN Ia according to the DD scenario. We also check for ONe WD and CO WD binary systems that merge in a Hubble time. Such a merger might lead to a peculiar SN Ia \citep{Kashyapetal2018}.

We assume that the main contributions to SNe Ia come from the DD scenario (sub-$M_{\rm Ch}$ explosions) and from the CD scenarios ($M_{\rm Ch}$ explosions). To identify binary WDs systems as potential progenitors of SNe Ia according to the DD scenario we require that they merge within a Hubble time of $10^{10} \yr$. We therefore compute for each binary WD system the merger time according to
gravitational waves (e.g., \citealt{Bertschinger2015})
\begin{equation}
t_{\rm GW}=\frac{5}{256}\frac{c^5}{G^3}\frac{r^4}{(m_1m_2)(m_1+m_2)},
\label{eq:tmerg}
\end{equation}
where $c$ is the speed of light, $G$ is the gravitational constant, $r$ is the separation between the two WDs at the end of the CE, and $m_1$ and $m_2$ are their masses. We assume that if the time until the merger is $t_{\rm GW}<10^{10} \yr$ and the two WDs are made of CO, then the system would become a type Ia SN in the frame of the DD scenario.

\section{RELATIVE CONTRIBUTIONS}
\label{sec:numbers}
\subsection{Evolutionary channels}
\label{subsec:evolution}
We run the population synthesis code that we describe in section \ref{sec:code} for four different values of the common envelope parameter, $\alpha_{\rm CE}=0.1, 0.3, 0.5$ and $1$.
We compare the relative contributions of seven binary outcomes, as listed in Tables
\ref{tableRelativeNoNS} and \ref{tableRelativeNS}. The total number of systems in these seven evolutionary channels is 532 for $\alpha_{\rm CE}=0.1$, 438 for $\alpha_{\rm CE}=0.3$, 487 for $\alpha_{\rm CE}=0.5$ and 701 for $\alpha_{\rm CE}=1.0$.
The statistical noise due to the finite number of systems is much smaller than some other uncertainties (section \ref{subsec:uncertianties}).
The first three outcomes are two close WDs that survived the CEE, but that will merge within a time of $10^{10} \yr$ by emitting gravitational waves, according to equation (\ref{eq:tmerg}).
The next four evolutionary routes end the CEE with a merger of the WD with the core of the AGB star. The remnant is either a massive WD, or a neutron star (NS). We consider only merger remnants with a mass of $M_{\rm rem} \ge 1.35 M_\odot$. We list these seven outcomes below according to the row numbers in Tables \ref{tableRelativeNoNS} and \ref{tableRelativeNS}.
\begin{enumerate}
  \item Two surviving CO WDs that merge by gravitational waves within $10^{10} \yr$ according to equation (\ref{eq:tmerg}). This late merger can lead to a SN Ia according to the DD scenario.
  \item Either the first WD to form is an ONe WD, or the remnant of the core of the AGB star in the final CEE is an ONe WD. The other is a CO WD. The merger of these two WDs might lead to a peculiar SN Ia (section \ref{subsec:peculiar}).
  \item The two surviving WDs are ONe rich. Their merger will most likely form a NS. The merger processes itself might release lots of energy, and might be a transient event. For example, radiation might come from an accretion disk or from jets.
  \item A CO WD merges with a CO core. If the remnant is a WD and not a NS, it might explode  after a delay and become a SN Ia according to the CD scenario.
  \item A CO WD merges with an ONe core. If the remnant is a WD it might explode as a peculiar SN Ia according to the CD scenario.
  \item An ONe WD merges with a CO core. If the remnant is a WD it might explode as a peculiar SN Ia according to the CD scenario.
   \item An ONe WD merges with an ONe core. The outcome is most likely a NS, and there will be no explosion.
\end{enumerate}
\begin{table*}[!htb]
\minipage{1\linewidth}
\centering
 \caption{Results from our population synthesis simulations for four values of the $\alpha_{\rm CE}$ parameter. We give the relative number of systems for each of the evolutionary channels described in section \ref{subsec:evolution} relative to the number of CO WD $+$ CO core mergers (channel 4). For channels 4 to 7 we list only merger products that are WDs with a mass of $M_{\rm WD} \ge 1.35 M_\odot$. }
    \begin{tabular}{| c | l | c | c | c | c |}
    \hline
    $N^{\rm o}$ & Evolutionary channel & $\alpha_{\rm CE}=0.1$ & $\alpha_{\rm CE}=0.3$ & $\alpha_{\rm CE}=0.5$ & $\alpha_{\rm CE}=1.0$  \\ \hline
    1 & DD: CO WD + CO WD            & 4.51     & 0.24     & 0.16     & 0.89     \\ \hline
    2 & DD: CO WD + ONe WD           & 0.08     & 0.14     & 0.21     & 0.76     \\ \hline
    3 & DD: ONe WD + ONe WD          & 0.60     & 0.10     & 0.11     & 0.26     \\ \hline
    4 & CD: CO WD + CO core (merger)   &  1       & 1        & 1        & 1        \\ \hline
    5 & CD: CO WD + ONe core (merger)  & $<$ 0.02 & $<$ 0.01 & $<$ 0.01 & $<$ 0.01 \\ \hline
    6 & CD: ONe WD + CO core (merger)  & 0.11     & 0.52     & 0.61     &  0.40    \\ \hline
    7 & CD: ONe WD + ONe core (merger) & $<$ 0.02 & $<$ 0.01 & $<$ 0.01 & $<$ 0.01 \\ \hline
    \end{tabular}
  	\label{tableRelativeNoNS}
\endminipage\hfill
\end{table*}
\begin{table*}[!htb]
\minipage{1\linewidth}
\centering
    \caption{Like Table \ref{tableRelativeNoNS} but now channels 4 to 7 also include the number of NSs. }
\begin{tabular}{| c | l | c | c | c | c |}
    \hline
    $N^{\rm o}$ & Evolutionary channel & $\alpha_{\rm CE}=0.1$ & $\alpha_{\rm CE}=0.3$ & $\alpha_{\rm CE}=0.5$ & $\alpha_{\rm CE}=1.0$  \\ \hline
    1 & DD: CO WD + CO WD      & 2.13 & 0.12 & 0.07 & 0.45 \\ \hline
    2 & DD: CO WD + ONe WD      & 0.04 & 0.07 & 0.10 & 0.38 \\ \hline
    3 & DD: ONe WD + ONe WD     & 0.29 & 0.05 & 0.05 & 0.13 \\ \hline
    4 & CD: CO WD + CO core (merger)   &  1   & 1    & 1    & 1    \\ \hline
    5 & CD: CO WD + ONe core (merger)  & 0.94 & 0.79 & 0.57 & 0.39 \\ \hline
    6 & CD: ONe WD + CO core (merger)  & 0.34 & 1.21 & 1.33 & 1.13 \\ \hline
    7 & CD: ONe WD + ONe core (merger) & 0.02 & 0.04 & 0.12 & 0.19 \\ \hline
    \end{tabular}
  	\label{tableRelativeNS}
\endminipage\hfill
\end{table*}

In Table \ref{tableRelativeNoNS} we list the relative numbers of WD systems in each channel relative to the number of systems in channel 4, that of a merger at the end of the CEE of CO WD with a CO core with a final remnant mass of $M_{\rm rem} \ge 1.35 M_\odot$. The last 4 rows include only systems that according to the population synthesis code we use are WD merger remnants.  In Table \ref{tableRelativeNS} we include also those mergers in channels 4-7 that end up as a NS.

\subsection{Major uncertainties}
\label{subsec:uncertianties}
The relative numbers of systems that end up in each evolutionary channel are quite sensitive to the value of $\alpha_{\rm CE}$, and in a complicated manner; see also \cite{Zhouetal2015} and \cite{Wangetal2017}. The reason is that this parameter influences the interaction at two phases: ($i$) when the initially more massive star becomes a giant and interact with the initially lower-mass star, which is still a main sequence star, and ($ii$) when the initially lower-mass star becomes a giant and interacts with the WD remnant of the initially more massive star.
Earlier population synthesis studies have examined the complicated matter of the parametrization of the CEE (e.g., \citealt{Toonenetal2012, Zhouetal2015, Wangetal2017}). We here simply take four values in the wide range of $\alpha_{\rm CE}=0.1-1$.
The fraction of channel 1, that might lead to SNe Ia in the DD scenario, to channel 4, that might lead to SNe Ia in the CD scenario, varies in the range of $0.16 \la (N_1/N_4) \la 4.5$
This shows that the uncertainties in population synthesis studies can be quite large.

The merger process introduces some uncertainties as well.
We consider the possibility that in some cases where our population synthesis code ends with a NS, a detailed treatment of the merger process would have ended in a rapidly rotating over-massive WD. The rotation delays the collapse of the WD (e.g., \citealt{Piersantietal2003, DiStefanoetal2011, Justham2011, Boshkayevetal2014, Wangetal2014, Benvenutoetal2015, MengHan2018}), that instead of a NS might give an explosion in some cases.
For that reason in Table \ref{tableRelativeNS} we list the relative number of systems, analogously as those listed in Table 1, but now including NS that are formed in each channel. A reasonable estimate should be in between these two tables, but closer to that of Table \ref{tableRelativeNoNS}.

Despite the large uncertainties that result from the usage of the $\alpha_{\rm CE}$ parameter, as well as other uncertainties that result from the poorly determined merger process, we do think we can make some important conclusions. These can be listed as follows.

\subsection{Comparing the DD and CD scenarios}
\label{subsec:DDCDscenarios}
The numbers of potential SN Ia progenitors in the DD scenario (channel 1) and in the CD scenario (channel 4) are generally comparable, $N_1 \equiv N_{\rm DD} \approx N_4 \equiv N_{\rm CD}$. This is in a general agreement with the finding that there are several times more observed double WD systems than is required by the DD scenario (e.g., \citealt{MaozHallakoun2017, Maozetal2018}), as well as several times the required number of massive WDs in the CD scenario \citep{BearSoker2018}.

\cite{Wangetal2017} conducted a thorough population synthesis study of the CD scenario, and found that the number of progenitors in the CD scenario and in the DD scenario are comparable, $N_{\rm CD} \approx N_{\rm DD}$, as we find here. They claim that the CD scenario might account for no more than $20 \%$ of all SNe Ia.
\cite{Zhouetal2015} found that for $\alpha_{\rm CE} \simeq 0.01-0.1$ the numbers are comparable, $N_{\rm CD} \approx N_{\rm DD}$, but for $\alpha_{\rm CE} \simeq 1$ not many systems are formed in the DD scenario, $N_{\rm CD} \gg N_{\rm DD}$.  The same authors claim that the CD scenario can account for about $2-10 \%$ of SNe Ia.
In an older paper \cite{MengYang2012} claim that the CD scenario can account for only $<1\%$ of all SNe Ia.

Our finding together with the existence of many potential SN Ia progenitors of the DD scenario (e.g., \citealt{MaozHallakoun2017, Maozetal2018}) and of the CD scenario \citep{BearSoker2018}, is compatible with the claim made by \cite{IlkovSoker2013} that the numbers of SN Ia progenitors in the CD scenario does not fall below that in the DD scenario and might account for most SNe Ia.

 In that regard, a word is in place here on the different conclusions of different population synthesis studies that we cited here, namely that some of these studies
conclude that the CD scenario can contribute at most few percents, and our claim that it can contribute more than half of all SNe Ia. We attribute the differences to the high sensitivity of the results to the common envelope treatment, in particular to the  $\alpha_{\rm CE}$ parameter  (section \ref{subsec:uncertianties}).
Naively, one would expect that when the value of $\alpha_{\rm CE}$ becomes smaller, namely, less efficient envelope removal, more WDs suffer merger with the core of the giant star, and the number of systems in the CD scenario increases at the expense of surviving binary WD systems, i.e. at the expense of systems in the DD scenario. But this is not so simple. Let us elaborate on this sensitivity.

As we discuss in section \ref{subsec:uncertianties} $\alpha_{\rm CE}$ influences the interaction at two phases. In the first phase the compact object that enters the envelope of the giant star is a main sequence star. In the second phase the compact object that enters the envelope of the giant star (that in the first phase was the main sequence star) is the WD that is the remnant of the core of the giant of the first phase. The reason that a decrease in the value of $\alpha_{\rm CE}$ not always increases the number of systems in the CD scenario is as follows. Although the lower value of $\alpha_{\rm CE}$ increases the probability of a WD to merge with the core, at the same time this lower value removes some systems from being potential SN Ia progenitors already in the first CEE phase.
But the efficiency of envelope removal should not be the same in the two phases. Indeed, \cite{Toonenetal2012} performed a population synthesis study where in the first phase they used the $\gamma$-formalism for the CEE. The $\gamma$-formalism is based on angular momentum considerations, rather than energy considerations in the $\alpha_{\rm CE}$ formalism.

 We suggest that future population synthesis studies use higher value of $\alpha_{\rm CE}$ in the first CEE phase. The reason is that main sequence companions are likely to launch jets just before and/or during the CEE, and these jets can efficiently remove envelop mass, and by that increasing the effective value of $\alpha_{\rm CE}$ (e.g. \citealt{Shiberetal2017, LopezCamaraetal2018}). We expect that a higher value of $\alpha_{\rm CE}$ in the first CEE phase will substantially increase the potential number of SN Ia progenitors in the CD scenario.

\subsection{Possible peculiar SNe Ia}
\label{subsec:peculiar}

It is possible that some massive WDs that contain both CO rich zones and ONe rich zones explode, but the explosion does not necessarily destroy the entire WD (e.g., \citealt{GilPonsGarciaBerro2001, Wangetal2014b, Kromeretal2015, Bravoetal2016,MengPodsiadlowski2018}). Such an explosion, or an outburst, might be observed as a peculiar SN Ia. We do not discuss here the outcome of the merger, and whether and how the CO and ONe-rich zones mix during the merger process. We only assume that some fraction, but not all, of the mergers of CO WD with an ONe core, or an ONe WD with a CO core, might lead to peculiar SNe Ia.

We take channel 2 to lead to some peculiar SNe Ia in the DD scenario, and channels 5 and 6 to lead to peculiar SNe Ia in the CD scenario. As we stated earlier, the contribution of each channel is between the numbers given in Table \ref{tableRelativeNoNS} and those given in Table \ref{tableRelativeNS}, but (much) closer to Table \ref{tableRelativeNoNS}.
Not all mergers lead to explosion whether in the CD or DD scenarios (as in many cases the remnant experiences an accretion induced collapse to a NS, e.g. \citealt{{WuWang2018}}). On the other hand, there are many other types of peculiar SNe Ia, like those that have helium (e.g., \citealt{Neunteufeletal2017, Toonenetal2018, Zenatietal2018}) and the failed-detonation supernova \citep{Jordanetal2012} that we do not treat at all here.

We simply conclude from Tables \ref{tableRelativeNoNS} and \ref{tableRelativeNS} that the number of peculiar SNe Ia that might result from the channels that we study here is a significant fraction of the number of regular SNe Ia. For example, compare row 6 to row 4 in Table \ref{tableRelativeNoNS}.

\section{BINARY PROPERTIES}
\label{sec:properties}

We describe here some properties of the systems that experience a merger during the CEE, focusing on WDs with masses above $1.35 M_\odot$ and NSs.
We only are going to discuss some properties that might be relevant to observations of SNe Ia and peculiar SNe Ia in the frame of the CD scenario. Consequently, we will not by any means describe the full evolution of the systems, and will not describe all properties.

In Fig. \ref{fig:menv03} we present the envelope mass of the giant star at the beginning of the CEE in the plane of the core of the giant star versus the mass of the WD that enters the envelope. Filled dots represent WDs with $M_{\rm WD} \ge 1.35 M_\odot$, and empty circles depict NSs. The most relevant property to take from this graph is that in many cases there is a massive envelope at the beginning of the CEE. This implies that if the explosion takes place within several hundreds thousand years after the merger then there will be a massive circumstellar matter. A short merger to explosion delay, $\la 10^6 \yr$, might take place for very massive WDs. Some of these massive WDs might also come from systems that the population synthesis code registers as NSs (section \ref{subsec:uncertianties}).
\begin{figure}[ht]
    \centering
    \includegraphics[width=0.5\textwidth]{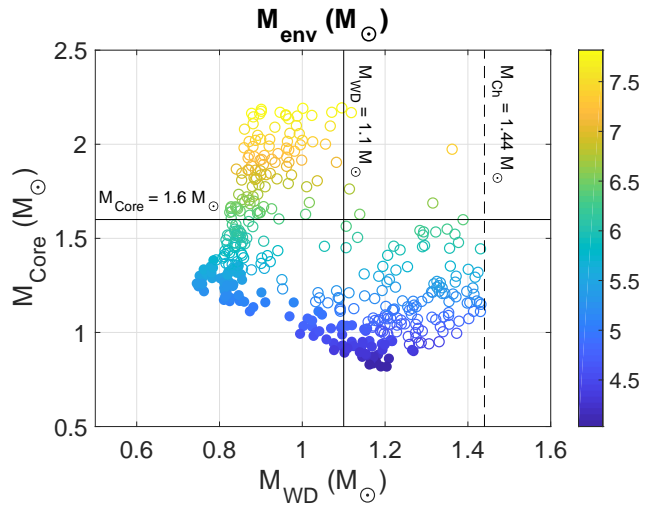}
    \caption{The mass of the envelope of the giant star at the beginning of the CEE for the population synthesis simulation with $\alpha_{\rm CE}=0.3$. The color bar is in units of $M_\odot$. The plane is the mass of the core of the giant star versus the mass of the WD that enters the envelope. Filled dots are WDs (all with $M_{\rm WD} \ga 1.35 M_\odot$) and hollow circles are NSs, after the CE is expelled. }
\label{fig:menv03}
\end{figure}

 Another interesting property is the initial semi-major axis of the binary system (when both stars are on the zero-age main sequence), which might significantly influence the course of the binary evolution. We present this quantity in Fig. \ref{fig:sep03}. We can see that there are two major branches of systems. Initially close systems that merge as a massive core and less massive WD, and systems with larger initial semi-major axis that merge when the core is about equal to or less massive than the WD. We do not get into the many complication details of the evolution, like for example the initial eccentricity or mass transfer episodes, given that the individual characteristics of each system are irrelevant when considering the ensemble properties of the whole population. The relevant point to our analysis is that the initial semi-major axis of many systems is less than about $2000 R_\odot$ ($\simeq 10 \AU$), implying that a tertiary star might be present at a stable orbital separation of $a_3 \la 100 \AU$.
 Such a tertiary star might stay bound even after relatively rapid mass loss episode of the CEE. Hence, a star might be left near the explosion site. This implies that the presence of a surviving star near the center of a supernova remnant, or near the explosion site of a SN Ia, does not necessarily point to the single degenerate scenario.
\begin{figure}[ht]
    \centering
    \includegraphics[width=0.5\textwidth]{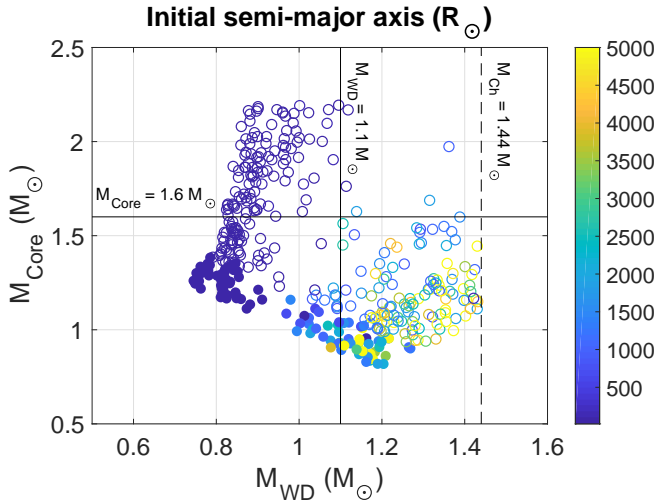}
    \caption{Like Fig. \ref{fig:menv03}, but now the color depicts the semimajor axis of the binary (when both stars were on the zero-age main sequence) in units of $R_\odot$.
For visual purposes, all system with initial semimajor axis above $5000 R_\odot$ have the same yellow color. }
    \label{fig:sep03}
\end{figure}

The population synthesis code calculates the masses of final merger remnants. But we note that the merger of two stars of a binary system is still a poorly understood mechanism, and the mass of the new star resulting from the coalescence of the two progenitors has many approaches, different from each other. Our code follows \cite{Hurley2002a} who use a relation between the final and initial binding energies of the envelope of the giant star to derive the final remnant mass (see their section 2.7.1).

In fig. \ref{fig:rem03} we show the mass of the remnant product. Many merger products that the population synthesis code returns as NS have masses below $1.4 M_\odot$. In this study we raise the possibility that some of these systems could instead be WDs, where rotation supports them against collapse. Some of these might eventually collapse to NS, and some might explode as SNe Ia or peculiar SNe Ia.
\begin{figure}[ht]
    \centering
    \includegraphics[width=0.5\textwidth]{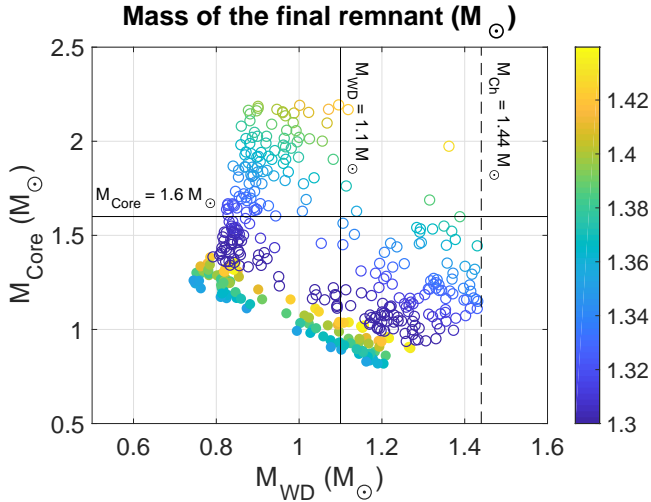}
    \caption{Like Fig. \ref{fig:menv03}, but now the color depicts the mass of the merger product in units of $M_\odot$. }
    \label{fig:rem03}
\end{figure}
\section{SUMMARY}
\label{sec:summary}

We used a population synthesis code (section \ref{sec:code}) to study the relative numbers of the different outcomes of a CEE of a WD with an AGB star. There are three evolutionary channels (numbers 1-3 in Tables \ref{tableRelativeNoNS} and \ref{tableRelativeNS}) that end with two WDs.
We examined only those binary WDs systems that merge within a time of $10^{10} \yr$ by emitting gravitational waves.
There are four evolutionary channels that end with a merger of the core of the AGB star and the WD during the CEE. In some cases a WD is formed, and in others a NS is formed. We considered only WD merger products that have a mass of $M_{\rm WD} \ge 1.35 M_\odot$, as we assume that some of these might explode as SNe Ia or peculiar SNe Ia.

In Table \ref{tableRelativeNoNS} we list the relative numbers of the outcome in each channel, and for four values of the common envelope parameter $\alpha_{\rm CE}$. Channels 4-7 include merger products that the code returns as WDs with $M_{\rm WD} \ge 1.35 M_\odot$. In Table
\ref{tableRelativeNS} channels 4-7 include also NS. The reason is that some of these remnants that the code registers as NSs have mass of less than about $1.4 M_\odot$ (Fig. \ref{fig:rem03}), and might be supported as a WD by rotation. Our study calls for simulations of the merger process of a CO core with an ONe WD and an ONe core with a CO WD, much as the simulation of CO core merger with a CO WD \citep{AznarSiguanetal2015}.

In Figs. \ref{fig:menv03}-\ref{fig:rem03} we present the envelope mass when the system enters the CEE, the initial semi-major axis, and the final remnant mass, respectively, for channels 4-7 and for $\alpha_{\rm CE} =0.3$.

From the tables we learn that the results of the population synthesis study are very sensitive to the common envelope parameter $\alpha_{\rm CE}$, as earlier studies have found  (e.g., \citealt{Toonenetal2012, Zhouetal2015, Wangetal2017}). Bearing in mind the uncertainties that this parameter, as well as the uncertainties of the merger product (section \ref{subsec:uncertianties}) introduce, we list here the main findings of our  population synthesis study.

(1) The number of merger products during the CEE that might explode as SNe Ia in the frame of the CD scenario (channel 4 in the tables), and the number of binary WDs that might lead to SNe Ia in the frame of the DD scenario (channel 1 in the tables) are of the same order of magnitude. That is, within the large uncertainties we found that $N_{\rm CD} \approx N_{\rm DD}$.

(2) We assume that the merger, during the CEE or later due to gravitational waves, of a CO WD (or a CO core) with an ONe WD (or an ONe core) might lead in some of the cases to peculiar SNe Ia.
Under this assumption our results show that the number of potential peculiar SNe Ia in the channels we studied here is non negligible, although less than the number of normal SNe Ia. There are other types of peculiar SNe Ia that we did not study here.

(3) As seen in Fig. \ref{fig:menv03} at the beginning of the CEE the envelope mass is quite large. This implies that if the explosion takes place within about $10^6 \yr$ form the CEE, whether in the CD scenario or in the DD scenario, or whether a normal or peculiar SN Ia, a massive circumstellar matter might be present at explosion time.

(4) Many systems that merge in the CD scenario have a relatively small initial semi-major axis, $\la 10 \AU$. Some of these systems might have a tertiary star in a stable orbit at a separation of about few hundreds AU or less that stays bound after the CEE. This implies that at explosion there might be a surviving star close to the explosion site, despite that the system did not come from the single degenerate scenario.
The surviving tertiary star, though, has some different properties than the companion in the single degenerate scenario. Firstly, the companion in our scenario is a main sequence star or a WD, while in the single degenerate scenario it is a main sequence star or a giant. In the double detonation scenario the companion can be a helium-rich WD. Secondly, at explosion the companion in our scenario is at a larger distance from the exploding WD than the companion in the single degenerate scenario or the double detonation scenario. Thirdly, as a result of that larger distance, after explosion the tertiary star will be ejected at a velocity of only several~$\km \s^{-1}$, whereas a main sequence companion in the single degenerate scenario or a WD companion in the double detonation scenario are ejected at velocities of $\ga 100 \km \s^{-1}$.

Overall, our results are compatible with, and support, the suggestion (table 1 in \citealt{Soker2018Rev}) that most $M_{\rm Ch}$ explosions, i.e., those with a WD progenitor of $M_{\rm WD} \simeq 1.4 M_\odot$, come from the CD scenario, and most  sub-$M_{\rm Ch}$ explosions come from the DD scenario.

\section*{Acknowledgments}
This research was initiated by Enrique Garc\'ia-Berro who sadly died in a tragic accident in September 2017. We express our deep appreciation to Enrique for enjoyable and enlightening collaboration over the years.  We thank an anonymous referee for useful comments.
This research was supported by the  Asher Fund for  Space Research at
the  Technion, and partially supported by the MINECO grant AYA\-2017-86274-P
and by the AGAUR (SGR-661/2017).


\label{lastpage}

\end{document}